\documentclass[a4paper,11pt]{article}
\pdfoutput=1 
\usepackage{jheppub}

\usepackage[T1]{fontenc} 
\usepackage[applemac]{inputenc}

\usepackage{amsmath, amssymb, dsfont}
\usepackage[numbers]{natbib}
\usepackage{graphicx}
\usepackage{enumerate}
\usepackage[normalem]{ulem}
\usepackage{empheq}

\newcommand{\rd}{\mathrm{d}}
\newcommand{\re}{\mathrm{e}}
\newcommand{\mixInd}[3]{{#1}^{#2}_{\hphantom{#2}#3}}
\newcommand{\comp}[2]{(\!\begin{smallmatrix}#1\\#2\end{smallmatrix}\!)} 

\title{\boldmath Fine-tuning with Brane-Localized Flux in 6D Supergravity}

\author[a,c]{Florian Niedermann}
\author[a,c]{and  Robert Schneider}

\affiliation[a]{Arnold Sommerfeld Center for Theoretical Physics, Ludwig-Maximilians-Universit\"at, Theresienstra{\ss}e 37, 80333 Munich, Germany}
\affiliation[c]{Excellence Cluster Universe, Boltzmannstra{\ss}e 2, 85748 Garching, Germany\\
~\\}
 
\emailAdd{florian.niedermann@physik.lmu.de}
\emailAdd{robert.bob.schneider@physik.uni-muenchen.de}

\abstract{
	There are claims in the literature that the cosmological constant problem could be solved in a braneworld model with two large (micron-sized) supersymmetric extra dimensions. The mechanism relies on two basic ingredients: First, the cosmological constant only curves the compact bulk geometry into a rugby shape while the 4D curvature stays flat. Second, a brane-localized flux term is introduced in order to circumvent Weinberg's fine-tuning argument, which otherwise enters here through a backdoor via the flux quantization condition. In this paper, we show that the latter mechanism does not work in the way it was designed: The only localized flux coupling that guarantees a flat on-brane geometry is one which preserves the scale invariance of the bulk theory. Consequently, Weinberg's argument applies, making a fine-tuning necessary again. The only remaining window of opportunity lies within scale invariance breaking brane couplings, for which the tuning could be avoided. Whether the corresponding 4D curvature could be kept under control and in agreement with the observed value will be answered in our companion paper~\cite{Niedermann:2015vbk}.
}

\begin{document}

\maketitle
\flushbottom


\section{Introduction and Summary}

Due to vacuum loops of massive particles, the value of the cosmological constant is not stable under radiative corrections and thus requires a tremendous amount of fine-tuning at all orders in perturbation theory. This simple observation gives rise to the famous cosmological constant (CC) problem~\cite{Weinberg:1988cp}. 

Braneworld scenarios provide an interesting territory in the search for a natural solution to the problem.  In particular, the special case of two codimensions has attracted a lot af attention because a 4D CC only leads to a conical singularity in extra space while the 4D brane geometry remains flat, like for a usual cosmic string. (A first discussion of the essential idea can be found in \cite{Rubakov:1983bz} and later adoptions in the context of large extra dimensions in~\cite{Chen:2000at,Carroll:2003db,Navarro:2003vw}). In other words, the gravitational impact of the CC is diverted from the intrinsic brane to the extrinsic bulk curvature.


One class\footnote{There is a separate class of models working with infinite extra dimensions. An additional, four dimensional, Einstein-Hilbert term localized on the brane is assumed to have a large enough coefficient to lead to an approximate 4D regime at the phenomenologically relevant scales. For the original proposal of the so-called \textit{brane induced gravity} model see~\cite{Dvali:2000hr} and for a more recent discussion in the 6D context also~\cite{Eglseer:2015xla}.} of models works with compact but large, viz.\ micron-sized, extra dimensions.\footnote{These models were first considered as candidates to solve the electroweak hierarchy problem~\cite{Antoniadis:1998ig,ArkaniHamed:1998nn}.} The compact space has the topology of a sphere and closes in two infinitely thin three-branes situated at the north and south pole. The sphere is stabilized by a nontrivial Maxwell flux wrapped around one of the compact dimensions. The standard model (SM) matter fields are localized on one of the two branes. In these models gravity obeys the four dimensional force law at large distances and becomes six dimensional at distances below the size of the compact dimension, i.e.\ the micrometer scale, which is still compatible with post-Cavendish experiments.\footnote{For recent experimental bounds see \cite{Kapner:2006si} and for a general review~\cite{Joyce:2014kja} (and references therein).}

Historically, these models evolved in three stages: First, the gravitational sector was described by six dimensional General Relativity (GR), cf.\ \cite{Chen:2000at,Leblond:2001xr,Carroll:2003db,Navarro:2003vw,Cline:2003ak}. In a second stage, the gravity sector was promoted to supergravity, thereby giving rise to the model of supersymmetric large extra dimensions (SLED) \cite{Aghababaie:2003wz}. On a technical level the model is extended by the dilaton as a new degree of freedom. The bulk theory is also (classically) scale invariant, which implies a flat dilaton direction in field space. As for the third and last stage, a brane-localized flux (BLF) term was added in the Maxwell sector of the theory~\cite{Burgess:2011mt, Burgess:2011va}. This terms allows to explicitly break scale invariance and, as a result, to lift the flat dilaton direction.

We mainly focus on the SLED model including the BLF term. Two questions are at the core of our work:

\begin{enumerate}
\item \textit{What is the condition for exact 4D flatness in this model?} It is clear that answering this question is of great interest with respect to the CC problem because we are looking for an explicit mechanism to hide the CC from a brane observer. Under the assumption of 4D maximal symmetry on the brane, we derive the desired condition in Sec.~\ref{sec:SLED}: \textit{Scale invariance is a sufficient condition for 4D flatness.} This is the central result of this work. In particular, the BLF term has to couple to the dilaton in a nontrivial way to preserve scale invariance. If scale invariance is broken, 4D flatness is not guaranteed; the actual value of the 4D curvature, however, can only be calculated in a regularized setup~\cite{Niedermann:2015vbk}.

While our result agrees with previous statements in the literature without BLF, it contradicts a former analysis with BLF term in \cite{Burgess:2011mt, Burgess:2011va}. There, it was claimed that regular, 4D flat solutions are ensured if the BLF term does not couple to the dilaton, which corresponds to an explicitly broken scale invariance. We discuss the origin of the mismatch.
\item \textit{Can the 4D flat solutions avoid Weinberg's no-go theorem~\cite{Weinberg:1988cp}?}
By reviewing the explicit solutions of the scale invariant model in Sec.~\ref{sec:explicit_sol}, we find a negative answer because these solutions require a tuning between the brane tension and other model parameters due to the flux quantization condition. In Sec.~\ref{sec:weinberg}, we argue that the reason can be traced back to scale invariance which entails a flat direction in field space. In particular, violating the tuning relation would lead to a run-away \`a la Weinberg.
\end{enumerate}

A key ingredient in our analysis is the addition of a counter term to the action, which is necessary (and sufficient) to dispose of divergences that arise due to the BLF and were missed in previous studies. This method, introduced in Sec.~\ref{sec:counter_term}, enables us to make quantitative predictions within the effective field theory (EFT) framework in which the branes are infinitely thin. An alternative approach was recently proposed in~\cite{Burgess:2015nka, Burgess:2015gba}, where the branes were microscopically resolved in a concrete UV model. We find perfect agreement with the results of~\cite{Burgess:2015nka}, as discussed in Appendix~\ref{ap:comp_UV}.

While---in a strict sense---our results only make a statement about the (non-)existence of a natural 4D Minkowski vacuum, they also raise concerns about the radiative stability of phenomenologically relevant de Sitter or quasi de Sitter solutions within the SLED model. However, the present work does not allow for a final statement about those solutions because, in general, the required scale-invariance breaking dilaton coupling leads to singularities at the brane positions. Therefore, this case requires some regularization, which is beyond the scope of the present paper and will be introduced and extensively discussed in a companion work~\cite{Niedermann:2015vbk}. We conclude with an assessment on the status of the model in Sec.~\ref{sec:conclusion}.


\subsection{Conventions and notation}

We use Weinberg's sign conventions~\cite{Weinberg_Gravitation}. Six dimensional spacetime coordinates are denoted by $ X^M $ ($ M = 0, \ldots, 5 $), four dimensional ones by $ x^{\mu} $ ($ \mu = 0, \ldots, 3 $), and the two extra space dimensions are labeled by $ y^m $ ($ m = 1, 2 $). Furthermore, $ \epsilon^{mn} $ is a tensor (not a density), i.e.~its components are $ \pm 1 / \sqrt{g_2} $. The delta function transforms as a density, so there is no metric determinant factor in its normalization condition $ \int\! \rd^2 y \, \delta^{(2)}(y) = 1 $.

\section{SLED with BLF} \label{sec:SLED}

The total action of the SLED model is given by~\cite{Burgess:2011mt}
\begin{equation} \label{eq:action}
	S = S_\mathrm{bulk} + S_\mathrm{branes} \,,
\end{equation}
where the bulk part is
\begin{equation} \label{eq:action_bulk}
	S_\mathrm{bulk} = - \int \rd^6 X \sqrt{-g}\,\left\{\frac{1}{2\kappa^2}\left[ R + (\partial_M\phi)(\partial^M\phi) \right] + \frac{1}{4} \re^{-\phi}F_{MN} F^{MN} + \frac{2e^2}{\kappa^4}\re^{\phi}\right\} ,
\end{equation}
and the brane contributions are
\begin{equation} \label{eq:action_brane}
	S_\mathrm{branes} = - \sum_b \int \rd^4 x \sqrt{-g_4} \left\{ \mathcal{T}_b(\phi)-\frac{1}{2} \mathcal{A}_b(\phi) \epsilon_{mn} F^{mn} \right\} .
\end{equation}
The field content consists of the 6D metric $ g_{MN} $ with corresponding Ricci scalar $ R $, a Maxwell field $ A_M $ with field strength $ F = \rd A $, and the dilaton $ \phi $. $ \kappa $ and $ e $ denote the gravitational and U(1)\footnote{This is not necessarily the U(1) gauged by $ A_M $, which can in general have a different coupling~\cite{Burgess:2011mt}, which we will denote by $ \tilde e $, and which appears in the flux quantization condition~\eqref{eq:flux_quant}.} coupling constants, respectively.
The sum in $ S_\mathrm{branes} $ runs over $ b \in \{ +, - \} $, corresponding to the two $ 3 $-branes at the north and south pole of the compact extra space. One of them should ultimately be identified with our universe. The first term $ \mathcal{T}_b $ is the brane tension (or 4D vacuum energy density), and the second term corresponds to the \emph{brane-localized flux} (BLF). At this stage, we allow both of them to have a priori arbitrary dilaton dependences. The central question we want to answer is how these must be chosen in order to obtain 4D flat solutions.

Under a (constant) rescaling $ g_{MN} \mapsto \zeta \, g_{MN} $, $ \re^{\phi} \mapsto \zeta^{-1} \, \re^{\phi} $, the bulk action transforms with a global scaling factor, $ S_\mathrm{bulk} \mapsto \zeta^2 S_\mathrm{bulk} $. This implies that the classical bulk equations of motion are scale invariant. This scale invariance is respected by the branes if
\begin{align} \label{scale_inv}
	\mathcal{T}_b(\phi) = \text{const}  && \text{and} && \mathcal{A}_b(\phi) \propto \re^{-\phi} \,,
\end{align}
and is broken otherwise.

\subsection{Ansatz}

We assume the geometry to be maximally symmetric in the four on-brane dimensions, as well as rotationally symmetric in extra space. This leads to the following most general ansatz,
\begin{subequations} \label{eq:ansatz}
\begin{align}
	\rd s^2 & = W^2(\rho) \,\hat g_{\mu\nu} \rd x^{\mu} \rd x^\nu + \rd \rho^2 + B^2(\rho) \rd \theta^2 \,, \label{eq:ansatz_met}\\
	A & = A_{\theta}(\rho)\rd\theta \,,\\
	\phi & = \phi(\rho) \,, \label{eq:ansatz_phi}
\end{align}
\end{subequations}
where the 4D metric $ \hat g_{\mu\nu} $ is maximally symmetric and thus completely characterized by the (constant) 4D Ricci scalar $ \hat R $.
The extra space is labeled by the azimuth angle $ \theta \in[0, 2\pi) $ and the polar angle $ \rho \in [\rho_+, \rho_-] $, with $ \rho_b = \rho_\pm $ denoting the brane positions at the north- and south pole, respectively, where~$ B|_{\rho=\rho_b} = 0 $.

\subsection{Maxwell Sector}

After inserting the identity $ 1 = \int \!\rd^2 y \, \delta^{(2)}(y) $ into the brane part~\eqref{eq:action_brane}, the total action~\eqref{eq:action} can be written as a single integral over a 6D Lagrangian density $ \mathcal{L} $. The Maxwell part is
\begin{equation}
	\mathcal{L}_{F} = -\sqrt{-g}\, \frac{1}{4}\re^{-\phi} F^2 + \frac{1}{2} \sum_b \sqrt{-g_4} \mathcal{A}_b(\phi)\,\epsilon_{mn}F^{mn}\,\delta^{(2)}(y-y_b) \,,
\end{equation}
where $ F^2 \equiv F_{MN}F^{MN} $, and the corresponding field equations read
\begin{equation} \label{Maxwell}
	\partial_M \left[\sqrt{-g}\, \re^{-\phi} F^{MN}-\sqrt{-g_4}\, \delta_m^M \delta_n^N \sum_b \mathcal{A}_b(\phi)\epsilon^{mn} \delta^{(2)}(y-y_b)\right] = 0 \;.
\end{equation}
With the ansatz~\eqref{eq:ansatz}, this gives the field strength\footnote{Note that in our conventions, and for the ansatz~\eqref{eq:ansatz}, $\epsilon^{\rho\theta} = 1 / B $, and $ \delta^{(2)}(y) = \delta(\rho) / (2\pi) $.}
\begin{equation} \label{Sol_Max}
	F_{\rho\theta}=\re^{\phi}\left[Q \, \frac{B}{W^4} + \frac{1}{2\pi} \sum_b \mathcal{A}_b(\phi) \delta(\rho - \rho_b) \right] ,
\end{equation}
where $Q$ is a constant of integration. However, this constant cannot be chosen freely, because flux quantization requires~\cite{Randjbar:1983, Burgess:2011va}
\begin{equation} \label{eq:flux_quant}
	Q \int \!\rd \rho \, \frac{\re^{\phi} B}{W^4} + \frac{1}{2\pi}\sum_b \left[ \mathcal{A}_b(\phi) \re^{\phi} \right]_{\rho=\rho_b} = \frac{n}{\tilde e} \qquad (n \in \mathds{N}) \,.
\end{equation}

We now encounter a peculiarity (which was missed in previous investigations~\cite{Burgess:2011mt, Burgess:2011va, Burgess:2013ara}): the presence of a localized delta-contribution to the field strength implies that the $ F^2 $ term in the action, which also enters the Einstein and dilaton equations, contains a divergent part $ \propto \delta(0) $. It is obviously caused by the BLF term, as it disappears for $ \mathcal{A}_b = 0 $, but is also definitely a relict of treating the branes as point-like.

At this point, there are two routes one can follow: Either, one gives up the idealization of infinitely thin branes and tries to come up with a UV model which microscopically resolves the branes. A first step in this direction was currently taken in~\cite{Burgess:2015nka, Burgess:2015gba}. Alternatively, one can ask if the divergence can be somehow removed, rendering the delta description possible again. In this work, we pursue the latter option. Physically speaking, it is motivated by the EFT expectation that all physical predictions should be insensitive to the microscopic details of an underlying UV model, as long as we ask low energy questions. In the case at hand, the CC problem manifests itself in the IR, at energies well below a realistic (inverse) brane thickness. Indeed, if it were necessary to understand the full UV physics in order to solve the CC problem, it would actually not solve the problem in the realm in which it is posed in the first place---as an IR problem in a low energy EFT.

There are several observations which give us further confidence that our approach captures the correct physical picture:
\begin{itemize}
\item All divergences can be completely removed by adding a single counter term to the action. 
\item In the special case of vanishing dilaton, where the concrete UV model~\cite{Burgess:2015nka} applies, our results are in perfect agreement with~\cite{Burgess:2015nka}, as will be discussed in Appendix~\ref{ap:comp_UV}.
\item In the end, it leads to the conclusion that $ \hat{R} = 0 $ is ensured for scale invariant brane couplings $ \mathcal{T}_b $ and $ \mathcal{A}_b $, in line with Weinberg's general arguments~\cite{Weinberg:1988cp}.
\end{itemize}

\subsection{Counter Term} \label{sec:counter_term}

Plugging the solution~\eqref{Sol_Max} back into the action yields
\begin{equation}
	S_F |_\mathrm{sol} = -\frac{1}{2} \int \rd^6 X \sqrt{-g}\, \re^{\phi} \frac{Q^2}{W^8} + S_\mathrm{div}\;,
\end{equation}
where the last term is the divergent contribution
\begin{equation}\label{eq:action_div}
	S_\mathrm{div} = \frac{1}{2} \sum_b \int \rd^4 x \sqrt{-g_4}\, \frac{\delta^{(2)}(0)}{\sqrt{g_2}} \re^\phi \mathcal{A}_b(\phi)^2\,.
\end{equation}
In order to arrive at a finite action, it is \emph{necessary} to introduce a counter term which cancels $ S_\mathrm{div} $, leading to the action
\begin{equation}\label{ren_action}
	\tilde S := S - S_\mathrm{div} \;.
\end{equation}
Below, we will see that this subtraction is also \emph{sufficient} in order to arrive at a consistent theory, because it removes all divergences from the Einstein and dilaton field equations. In other words, the theory defined by the action $ \tilde S $ provides an explicit realization of the SLED model with a BLF term, which still allows for consistently treating the branes as infinitely thin---unlike for the original action $ S $.

Several further comments regarding the counter term $ S_\mathrm{div} $ are in order:
\begin{itemize}
	\item Since it does not contain $ A_M $, the Maxwell equations~\eqref{Maxwell} and the corresponding solution~\eqref{Sol_Max} are not affected. (Otherwise, it could have been necessary to reiterate the process and introduce further counter terms.)
	
	\item It has the correct symmetries to qualify as a legitimate 4D brane action because the combination $ \delta^{(2)}(y) / \sqrt{g_2} $ is a scalar.

	\item For later reference, note that the term preserves the scale invariance of the theory for the choice $\mathcal{A}_{b}(\phi)\propto \re^{-\phi}$.
	
	\item It cannot be viewed as a renormalization of the brane tension because of the factor $ 1 / \sqrt{g_2} $. Due to this explicit dependence on the bulk metric, it will enter the Einstein equations differently than $ \mathcal{T}_b(\phi) $, see~\eqref{eq:einstein_full}. But this factor is dictated both by general covariance, and the requirement to successfully cancel all divergences.
	
	\item The ill-defined quantity $ \delta^{(2)}(0) $ should better be thought of as a large but finite constant, as would arise in some actual regularization, where the delta function is replaced by some smeared function\footnote{For concreteness, in our coordinates one could consider $ \Theta(\epsilon - \rho) \rho / (\pi \epsilon^2) $. However, it is not quite clear how the BLF term could be modeled in such a regularization. A more consistent regularization was recently proposed in~\cite{Burgess:2015nka, Burgess:2015gba}, which gives results in full agreement with our predictions, see Appendix~\ref{ap:comp_UV}.} which has support in a small region of proper radius $ \epsilon $. In this case, $ \delta^{(2)}(0) / \sqrt{g_2} $ would be replaced by $ \sim 1 / \epsilon^2 $.
\end{itemize}

In the following, we present the dilaton and Einstein equations which are obtained from the action $ \tilde S $.

\subsection{Dilaton Sector} \label{sec:dilaton}

The dilaton equation is
\begin{multline} \label{eq:dilaton_eom}
	\frac{1}{\kappa^2}\Box \phi + \frac{1}{4}\re^{-\phi} F^2 - \frac{2e^2}{\kappa^4} \re^{\phi}\\
+ \sum_b \frac{\delta^{(2)}_b}{\sqrt{g_2}} \left\{-\mathcal{T}^{\prime}_b+\frac{1}{2}\mathcal{A}^{\prime}_b\, \epsilon_{mn} F^{mn} - \frac{\delta^{(2)}(0)}{\sqrt{g_2}} \, \re^{\phi} \mathcal{A}_b\left[\frac{1}{2}\mathcal{A}_b+ \mathcal{A}^{\prime}_b  \right] \right\} = 0 \;,
\end{multline}
where the primes here denote $ \rd / \rd\phi $, and $ \delta^{(2)}_b \equiv \delta^{(2)}(y - y_b) $.
The last term, proportional to $\delta^{(2)}(0)$, follows from the counter term in \eqref{ren_action}. Once we substitute the Maxwell solution \eqref{Sol_Max}, all divergent contributions exactly cancel as advertised, and the dilaton equation becomes, for the ansatz~\eqref{eq:ansatz},
\begin{equation}\label{Dilaton}
\frac{1}{\kappa^2}\Delta_2 \phi + \frac{1}{2}\re^{\phi}\left( \frac{Q^2}{W^8} - \frac{4e^2}{\kappa^4} \right) 
+ \sum_b \frac{\delta^{(2)}_b}{B} \left\{ Q \frac{\re^{\phi}}{W^4} \left(\mathcal{A}^{\prime}_b+\mathcal{A}_b\right)-\mathcal{T}^{\prime}_b \right\} = 0 \;,
\end{equation}
where the covariant 2D Laplace operator is\footnote{A prime denotes derivative with respect to the argument of the function: for a $ \rho $-dependent function (like $ \phi $) the prime is $ \rd / \rd\rho $, for the brane couplings (like $ \mathcal{T}_b $) it still denotes $ \rd / \rd\phi $.}
\begin{equation} \label{eq:laplace_phi}
	\Delta_2 \phi = \frac{1}{BW^4} \left ( BW^4 \phi' \right )' = \phi'' +  \left (\frac{B'}{B} + \frac{4W'}{W} \right) \phi' \,.
\end{equation}

We now integrate this equation over a small $\epsilon$-disc covering either the north or the south pole. By using Stokes' theorem and taking the limit $\epsilon \rightarrow0$, we find the following boundary condition
\begin{equation} \label{eq:dilaton_bc}
	\left[B \phi' \right]_{\rho=\rho_b} = - \frac{\kappa^2}{2\pi} \left[ Q \frac{\re^{\phi}}{W^4} \left(\mathcal{A}^{\prime}_b + \mathcal{A}_b\right) - \mathcal{T}^{\prime}_b \right]_{\rho=\rho_b}\,.
\end{equation}
In the scale invariant case~\eqref{scale_inv}, the right hand side of the last equation is zero, which in turn allows for a regular dilaton profile with vanishing $\rho$-derivatives at both brane positions. On the other hand, if the right hand side is nonzero (as expected in the non scale invariant case), the $ \phi $ profile becomes singular (thereby also implying a curvature singularity) since $ B \to 0 $ at the branes. As a more explicit study of this case requires to regularize the setup, the reader is referred to our companion work~\cite{Niedermann:2015vbk} where the non scale invariant case is investigated in a thick brane model.


\subsection{Gravitational Sector}

The Einstein equations read
\begin{multline} \label{eq:einstein_full}
	\frac{1}{\kappa^2} \left[ \mixInd{G}{M}{N} + \left (\partial^M\phi\right )\left (\partial_N\phi\right ) - \frac{1}{2} \delta^M_N (\partial\phi)^2 \right] + \re^{-\phi} F^{MP}F_{NP} -  \delta^M_N \left[ \frac{1}{4} \re^{-\phi} F^2 + \frac{2e^2}{\kappa^4} \re^{\phi} \right] \\
	= \sum_b \frac{\delta^{(2)}_b}{\sqrt{g_2}} \left\{ \delta^M_\mu \delta^\mu_N \, \mathcal{T}_b  + \left( \delta^M_m \delta^m_N - \delta^M_\mu \delta^\mu_N\right) \frac{\mathcal{A}_b}{2} \left[  \epsilon_{mn} F^{mn} - \re^\phi \mathcal{A}_b \frac{\delta^{(2)}(0)}{\sqrt{g_2}} \right] \right\} ,
\end{multline}
where $ (\partial\phi)^2 \equiv \left (\partial_M \phi\right ) \left (\partial^M \phi\right ) $. After plugging in the solution for the Maxwell field, again all the $ \delta^{(2)}(0) $-terms cancel.
For the ansatz~\eqref{eq:ansatz}, there are three nontrivial Einstein equations---the $ \comp{\mu}{\nu} $, $ \comp{\rho}{\rho} $ and $ \comp{\theta}{\theta} $ components---which explicitly read
\begin{subequations} \label{eq:einstein_expl}
\begin{align}
	- \frac{1}{\kappa^2} \left ( \frac{\hat R}{4 W^2} + 3 \frac{W''}{W} + \frac{B''}{B} + 3\frac{W'^2}{W^2} + 3 \frac{W'B'}{WB} + \frac{1}{2} \phi'^2  \right ) & =\frac{\re^\phi}{2} \left( \frac{Q^2}{W^8} + \frac{4e^2}{\kappa^4} \right) + \sum_b \frac{\delta^{(2)}_b}{B} \mathcal{T}_b \,, \\
	\frac{1}{\kappa^2} \left ( \frac{\hat{R}}{2W^2} + 6\frac{W'^2}{W^2} + 4\frac{W'B'}{WB} - \frac{1}{2} \phi'^2 \right ) & = \frac{\re^{\phi}}{2} \left( \frac{Q^2}{W^8} - \frac{4e^2}{\kappa^4} \right) \,, \label{eq:einstein_rho}\\
	\frac{1}{\kappa^2} \left ( \frac{\hat{R}}{2W^2} + 4\frac{W''}{W} + 6\frac{W'^2}{W^2} + \frac{1}{2} \phi'^2  \right ) & = \frac{\re^{\phi}}{2} \left( \frac{Q^2}{W^8} - \frac{4e^2}{\kappa^4} \right) \,, \label{eq:einstein_theta}
\end{align}
\end{subequations}
respectively.
The difference of the $ \comp{\rho}{\rho} $ and $ \comp{\theta}{\theta} $ equations is
\begin{align}\label{eq:warping}
	\frac{W''}{W} - \frac{W'B'}{WB} + \frac{1}{4} \phi'^2 & = 0 \,,
\end{align}
which shows that a nontrivial dilaton profile necessarily implies a warped geometry. Thus, it was necessary to include the warping factor $W$ in \eqref{eq:ansatz_met} in order to allow for generic statements about the 4D maximally symmetric setup.

\subsection{Condition for 4D Flatness}

We now want to answer the question how the brane couplings $ \mathcal{T}_b(\phi) $ and $ \mathcal{A}_b(\phi) $ must be chosen in order to obtain 4D flat solutions. To this end, we consider the 2D trace of the Einstein equations, i.e.~the sum of~\eqref{eq:einstein_rho} and~\eqref{eq:einstein_theta}, which gives
\begin{equation} \label{eq:einstein_2D_trace}
	\frac{1}{\kappa^2} \left( \frac{\hat R}{W^2} + 4 \Delta_2 \ln W \right) = \re^{\phi} \left ( \frac{Q^2}{W^8} - \frac{4e^2}{\kappa^4} \right ) \,,
\end{equation}
with $ \Delta_2 $ defined as in~\eqref{eq:laplace_phi}. Using the dilaton equation~\eqref{Dilaton}, we can rewrite this as
\begin{equation}
	\frac{1}{2 \kappa^2} \left[ \frac{\hat R}{W^2} + 2 \Delta_2 \left ( \phi + 2 \ln W \right ) \right] = -\sum_b \frac{\delta^{(2)}_b}{B} \left[ Q \frac{\re^{\phi}}{W^4} \left(\mathcal{A}^{\prime}_b + \mathcal{A}_b \right) - \mathcal{T}^{\prime}_b \right] \,,
\end{equation}
Following~\cite{Gibbons:2003di}, we multiply this equation with $ BW^4 $ and integrate over the whole extra space, yielding
\begin{empheq}[box=\fbox]{equation} \label{eq:degrav_cond}
	\hat R  = - \frac{2 \kappa^2}{V} \sum_b  \left[ Q \re^\phi \left(\mathcal{A}^{\prime}_b + \mathcal{A}_b\right) - W^4 \, \mathcal{T}^{\prime}_b\, \right]_{\rho = \rho_b} \,,
\end{empheq}
with
\begin{equation}
	\label{def:volume}
	V := 2\pi \int \!\rd\rho\, BW^2 = \int \rd^2 y \, \sqrt{g_2} \, W^2 \,.
\end{equation}
Equation~\eqref{eq:degrav_cond} is the central result of our analysis, relating the on-brane curvature $ \hat R $ to the brane couplings. The only assumption necessary for its derivation was to have a 4D maximally symmetric geometry, allowing for the ansatz~\eqref{eq:ansatz}. Most importantly, it shows that 4D flatness is guaranteed by scale invariant dilaton-brane couplings~\eqref{scale_inv}, and \emph{not} by dilaton independent couplings ($ \mathcal{A}'_b = \mathcal{T}'_b = 0 $) in the presence of a BLF term as was previously claimed in the literature~\cite{Burgess:2011mt, Burgess:2011va, Burgess:2013ara}. If scale invariance is broken, the right hand side of~\eqref{eq:degrav_cond} does not vanish identically; however, since it explicitly depends on $ \phi_b $---which can generically diverge in this case---an actual evaluation requires to regularize the setup~\cite{Niedermann:2015vbk}.


Alternatively, using the dilaton boundary condition~\eqref{eq:dilaton_bc}, Eq.~\eqref{eq:degrav_cond} can be written as
\begin{equation}
	\hat R  = \frac{4\pi}{V} \sum_b \left[BW^4 \phi' \right]_{\rho=\rho_b} \,,
\end{equation}
saying that a necessary and sufficient condition for 4D flatness is a regular dilaton profile at the branes. This was already observed in~\cite{Aghababaie:2003wz}. But in \cite{Burgess:2011mt, Burgess:2011va, Burgess:2013ara}, the wrong conclusion was drawn that this would be equivalent to dilaton independent brane couplings ($ \mathcal{A}'_b = \mathcal{T}'_b = 0 $).\footnote{The error was caused by using the dilaton boundary condition from Ref.~\cite{Bayntun:2009im}, which is only applicable in the case without BLF.} This is not the correct condition, because the BLF term leads to an additional, indirect dilaton coupling. This was explicitly shown in Sec.~\ref{sec:dilaton}: Due to the bulk $ F^2 $ term  the dilaton equation~\eqref{eq:dilaton_eom} obtains an $ \delta^{(2)} $-contribution proportional to $\mathcal{A}_b$ (in addition to the $\mathcal{A}'_b$ term). In other words, even if there is no \emph{direct} dilaton-brane coupling, $ \phi $ still gets sourced \emph{indirectly} by the BLF term, because it couples to the bulk Maxwell field. Instead, we have proven that it is scale invariance which ensures the brane dimensions to remain flat despite the presence of a brane vacuum energy.

\section{Explicit 4D Flat Solutions}\label{sec:explicit_sol}

Let us now specialize to the case of 4D flat solutions, i.e.~$ \hat R = 0 $, which are the relevant candidates with respect to the CC problem. As shown above, this is guaranteed by brane-dilaton couplings of the form
\begin{align}
	\mathcal{T}_b = \text{const} \,, && 	\mathcal{A}_b = \Phi_b \,\re^{-\phi} \,,
\end{align}
which preserve the scale invariance of the bulk theory. As we have seen, this also implies a regular dilaton profile, see Eq.~\eqref{eq:dilaton_bc}. Incidentally, the most general solutions are explicitly known for this setup~\cite{Gibbons:2003di}:\footnote{We use the coordinates introduced in~\cite{Burgess:2013eua}. The metric could as well be brought into the form~\eqref{eq:ansatz_met} by changing to the normal coordinate $ \rho $ via $ \rd \rho \propto W \rd \xi $, but this transformation yields complicated expressions containing hypergeometric functions, which are not very useful.}
\begin{subequations}
\begin{align}
	\rd s^2 & = W^2(\xi) \left[ \eta_{\mu\nu} \rd x^\mu \rd x^\nu + \re^{-\phi_0} r_B^2 \left ( \rd \xi^2 + \frac{\alpha_+ \alpha_-}{W^8(\xi)} \sin^2(\xi) \rd \theta^2 \right ) \right] \,,\\
	\phi(\xi) & = \phi_0 - 2\ln W(\xi) \,, 
	\intertext{with}
	W^4(\xi) & = \cosh(v) - \sinh(v) \cos(\xi) \,.
\end{align}
\end{subequations}
The constants $ r_B, \alpha_\pm $ and $ v $ are fixed in terms of the model parameters via
\begin{align}
	\label{def:param}
	r_B = \frac{\kappa}{2e} \,, && \alpha_\pm = 1 - \frac{\kappa^2}{2\pi} \mathcal{T}_\pm \,, && v = \frac{1}{2} \ln \left ( \frac{\alpha_+}{\alpha_-} \right ) \,,
\end{align}
with $ \pm $ labeling the two branes, located at the poles $ \xi_+ = 0 $ and $ \xi_- = \pi $. The dilaton constant $ \phi_0 $ is not determined by any of the field equations, as is guaranteed by scale invariance. Geometrically, the parameters $ 2\pi(1 - \alpha_\pm) $ correspond to the deficit angles at the branes that are created by their tensions. In the special case of equal tensions, the solution simplifies to the rugby ball geometry. Otherwise, the warping $ W $ is nontrivial and the extra space looks like a deformed rugby ball. 
The extra space volume $ V $ is
\begin{equation}
	V = 4\pi r_B^2 \sqrt{\alpha_+\alpha_-} \,\re^{-\phi_0} \,.
\end{equation}
Furthermore, in these coordinates, the Maxwell field strength is given by
\begin{equation}
	F_{\xi\theta} = \frac{r_B}{\kappa} \sqrt{\alpha_+ \alpha_-} \, \frac{\sin(\xi)}{W^8(\xi)} + \frac{1}{2\pi} \sum_b \Phi_b \,\delta(\xi - \xi_b) \,.
\end{equation}
The flux quantization condition~\eqref{eq:flux_quant} then becomes
\begin{empheq}[box=\fbox]{equation}
	\label{eq:flux_quant_2}
	\frac{\sqrt{\alpha_+ \alpha_-}}{e} + \frac{1}{2\pi} \sum_b \Phi_b = \frac{n}{\tilde e} \,.
\end{empheq}

\section{Weinberg's No-Go Theorem}\label{sec:weinberg}
The above results allow us to further clarify the status of the SLED model as a potential solution to the CC problem. To be specific, we ask whether this particular 6D construction can avoid Weinberg's no-go theorem~\cite{Weinberg:1988cp} in 4D. The original idea was to look for a classical \textit{adjustment mechanism}\footnote{The mechanism is also dubbed \textit{self-tuning} in the literature.} that is able to entirely prevent a CC from gravitating. Without going into any details, the theorem can be summarized in two main statements:

\begin{itemize}
	\item A dynamical adjustment to zero curvature demands the existence of a scalar potential of the form 
	\begin{align}
		V(\varphi,\sigma)= \re^{-\varphi} V_0(\sigma)\,,
	\end{align}
	 where the field $\sigma$ is allowed to be massive. The potential $V_0(\sigma)$ is required to vanish at its minimum. Then, $\varphi$ corresponds to a flat direction in field space and can be understood as the Goldstone boson of a spontaneously broken scale invariance of some parent theory. 
		
	\item The condition $V_0|_{\rm min}=0$ is a tuning relation on model parameters. In fact, quantum corrections generically lift the minimum, thus implying a run-away behavior of $\varphi$ towards infinity. This is at odds with both having a vanishing curvature as well as a theory with massive particles. 
\end{itemize} 
So the outcome is quite sobering: Either we are back to the original tuning relation, or we encounter a run-away behavior.

In order to answer the initiatory question, we will follow a historical approach: First, we discuss the shortcomings of two predecessor models: The non-SUSY model, i.e.\ the pure rugby model in the context of 6D GR, as well as the simplest supersymmetric model, which here corresponds to adding solely the dilaton.  Second, we discuss the full-fledged SLED model with BLF term.

\subsection{Towards SLED}

The simplest model, as discussed in \cite{Chen:2000at,Carroll:2003db,Navarro:2003vw,Cline:2003ak}, is based on six dimensional GR on a compact manifold closing up in two pure tension branes. The whole system is stabilized by a non-vanishing Maxwell flux that is wrapped around one of the compact directions. Since this case was discussed extensively in the literature, we limit ourselves to a brief summary of the main obstacle. It arises due to the flux quantization\footnote{Without quantization, there would also still be the problem of flux \emph{conservation}, which would forbid a change in tension, as occurs during a phase transition~\cite{Garriga:2004tq}.} condition \eqref{eq:flux_quant} which imposes a parameter constraint on the brane tension,\footnote{For simplicity, we limited ourselves to the case with two identical pure tension branes corresponding to a geometry without warping.} see \cite{Navarro:2003bf,Nilles:2003km,Garriga:2004tq} and for a recent review on the topic also \cite{Burgess:2013ara},
\begin{equation}
	\alpha^2 = \frac{n^2\,\kappa^4}{2e^2} \Lambda \;,
\end{equation}
where $\Lambda$ is the bulk CC, as expected to be present in the non supersymmetric case.
A violation of this relation would lead to a non-vanishing 4D curvature. In particular, the system is not able to readjust after a change in the brane tension, as would occur during a phase transition. Thus, the tension has again to be tuned in order to achieve the desired value that is compatible with observations. To put it differently, the tuning problem is as severe as it was in standard 4D GR.

%
%

The situation in the supersymmetric model \eqref{eq:action} without BLF is not any better. Although the dilaton shows up as an additional degree of freedom in this case, the tuning cannot be avoided. In fact, it only changes its guise and now manifests itself as Eq.\ \eqref{eq:flux_quant_2} (without BLF term), which can be read as a new tuning relation on the brane tensions.


The origin of this tuning problem is linked to the fact that $ \phi_0 $ is not determined by any of the field equations. As already mentioned, this is due to the scale invariance of the theory. Therefore, $ \phi_0 $ does not appear in the flux quantization condition, which thereby imposes a tuning relation on the model parameters. If this tuning is satisfied, there are 4D flat solutions for any value of $ \phi_0 $, corresponding to a flat dilaton potential; otherwise, there would be a run-away behavior \`a la Weinberg. This follows from the observation that the GGP solutions presented in Sec.~\ref{sec:explicit_sol} are the most general 4D maximally symmetric solutions in the (scale-invariant) pure tension case~\cite{Gibbons:2003di}.

\subsection{SLED with BLF}

The crucial idea of \cite{Aghababaie:2003wz,Burgess:2011mt} to ameliorate the situation was to introduce the BLF terms in the brane action \eqref{eq:action_brane}. In general, these terms break scale invariance explicitly; thus the hope was to lift the flat dilaton direction, which is caused by scale invariance, thereby avoiding the degeneracy in field space. To be more precise, from \eqref{eq:flux_quant} it is clear that in the case with broken scale invariance, the dilaton does not drop out of the flux quantization condition. Thus, instead of being a tuning relation on the brane tension, this equation is expected to simply fix the value of $\phi_0$. Now, the crucial observation is that according to our central Eq.~\eqref{eq:degrav_cond}, an explicitly broken scale invariance---while avoiding the tuning issue---does not imply a flat 4D vacuum (as was previously claimed).

On the other hand, when we consider the scale invariant setup, the fine-tuning issue gets restored via Eq.~\eqref{eq:flux_quant_2} because the dilaton drops out. As before, since the GGP solutions are the most general static solutions of the scale invariant setup, it is clear that violating the tuning relation has to lead to some kind of run-away behavior, in complete agreement with Weinberg's argument.\footnote{Note that we did not formulate the tuning problem in terms of an explicit 4D potential as done by Weinberg. This could be achieved by performing a KK reduction as in~\cite{Burgess:2015lda}}

\section{Conclusion}\label{sec:conclusion}

The logic of previous works was based on a two-step approach to the CC problem: 

\begin{enumerate}
	\item
	Prove the existence of a flat 4D solution in the presence of a brane tension without implicitly imposing a fine-tuning of model parameters (which would be radiatively unstable). 
	\item
	Find a 4D (quasi) de Sitter solution with a small effective CC which is radiatively stable and compatible with the observed value.
\end{enumerate}

It is clear that once the first task is accomplished, the prospects for the second one are rather positive: it is quite conceivable that it can be achieved by slightly deforming the perfectly flat solutions. However, our results show that the first task is already not realizable within the SLED model. While this does not necessarily imply that the second---and also crucial---one is bound to fail, it at least means that the outcome is completely open again.

Equation \eqref{eq:degrav_cond} shows that in order to address the second task, one has to break scale invariance explicitly. Then, due to \eqref{eq:dilaton_bc}, the dilaton derivative at the brane position does not vanish. In general, the dilaton will diverge at the brane position, thereby also implying a curvature singularity mediated by the warping factor $W$ according to \eqref{eq:warping}. Deriving the solutions of the full bulk-brane system in the non scale invariant case thus requires the use of an appropriate regularization of the brane (or even a UV model). This is beyond the scope of the present work but successfully realized in a companion paper~\cite{Niedermann:2015vbk}. 

To conclude, we have shown that it is scale invariance which ensures the brane dimensions to remain flat, provided the model parameters satisfy a tuning relation. While these findings cast doubts on the model's ability to address the CC problem, a further discussion of the scale invariance breaking case will lead to a final verdict.

\subsection*{Note Added}
The preprint \cite{Burgess:2015gba} appeared shortly before our posting. There, instead of a delta-analysis, an explicit vortex construction is used to describe the brane sector. In agreement with our results the authors find that scale invariance ensures a vanishing 4D curvature.

Moreover, in a recent comment~\cite{Burgess:2015kda} on our work it was argued that the radial Einstein constraint~\eqref{eq:einstein_rho} is at odds with non scale invariant brane couplings. The conclusion of~\cite{Burgess:2015kda} was that one has to take into account an ad hoc metric dependence of the delta function, designed such that it leads to a localized contribution to the angular Einstein equation~\eqref{eq:einstein_theta}; in this case the constraint fixes the size of this term, and non scale invariant couplings are possible.
While we discuss this issue in more detail in our companion paper~\cite{Niedermann:2015vbk}, let us summarize here:
\begin{enumerate}[(i)]
	\item Physically, the additional term corresponds to an angular pressure component. Since a codimension-two brane contains no direction this pressure could act in, such a construction seems ill-defined.
	
	\item If one still accepts this term, none of the conclusions of this work would change. There would be an additional contribution to $ \hat{R} $  in~\eqref{eq:degrav_cond} which also vanishes in the scale invariant case.
	
	\item Alternatively, the impossibility to break scale invariance consistently on a delta brane in the case of 4D maximal symmetry could be regarded as a prediction of this analysis. (Note that this does not preclude scale invariance breaking couplings; scale invariance could in principle also be restored dynamically by a runaway behavior.)
	
	\item By studying a regularized setup and properly taking the thin brane limit in~\cite{Niedermann:2015vbk}, we will show that the latter option is indeed realized for a relevant class of couplings. While this means that $ \hat{R} \to 0 $ despite these non scale invariant couplings, it does not save the model because either one has to tune certain parameters or phenomenological bounds are violated.
\end{enumerate}


\acknowledgments
We thank Cliff Burgess, Ross Diener, Stefan Hofmann, Tehseen Rug and Matthew Williams for helpful discussions. We are in particular thankful to Stefan Hofmann for encouraging us to publish this work. We thank the Perimeter Institute of Theoretical Physics for its hospitality during an early stage of this work.
The work of FN was supported by TRR 33 ``The Dark Universe''.
The work of RS was supported by the DFG cluster of excellence ``Origin and Structure of the Universe''.


\appendix
\section{Agreement with a Specific UV Model} \label{ap:comp_UV}

A crucial step in our analysis was the introduction of the counter term in Sec.~\ref{sec:counter_term}, which was necessary to handle delta-like branes in the presence of BLF terms. As argued in Sec.~\ref{sec:counter_term}, this term is uniquely fixed in the sense that it is necessary and sufficient to remove all divergences, and has all the required symmetries. However, in order to gain more confidence in our approach, it is instructive to compare our results with the ones obtained in a recently proposed UV model~\cite{Burgess:2015nka}. There, the microscopic degrees of freedom creating the branes are resolved, and the branes have a finite thickness, thus avoiding all divergences. The BLF is modeled by introducing a kinetic mixing of the Maxwell field to another U(1) field. In~\cite{Burgess:2015nka}, only the non-supersymmetric case (without dilaton) was studied. This case is also covered by our analysis, and can be recovered by setting the dilaton to zero ($ \phi \equiv 0 $), discarding its equation of motion, Sec.~\ref{sec:dilaton}, and replacing $ 2e^2 / \kappa^4 $ by a bulk CC $ \Lambda $.

The main result of~\cite{Burgess:2015nka} was that  the BLF ``does not gravitate''. This result is recovered in our analysis by noticing that $ \mathcal{A}_b $ does not appear in the Einstein equations~\eqref{eq:einstein_expl}. Note that in the original equation~\eqref{eq:einstein_full} it \emph{does} appear; it only drops out after plugging in the solution for the Maxwell field, i.e.~it is canceled by the localized contributions from $ F_{MN} $. This is exactly the cancellation mechanism which is describes in the paragraph below Eq.~(3.71) in~\cite{Burgess:2015nka}. In summary, there is no contribution of the localized flux to the 4D Ricci on the brane in the non-SUSY case.


Reference~\cite{Burgess:2015nka} also discusses how the UV results can be understood in a low energy EFT in which the branes look delta-like. It is also found that a renormalization is necessary to avoid divergent terms in the 4D action. Remarkably, this renormalization is \emph{exactly} the same subtraction scheme we suggested here in Sec.~\ref{sec:counter_term}.\footnote{Let us emphasize, though, that our technique and results were communicated to the authors of~\cite{Burgess:2015nka} long before~\cite{Burgess:2015nka} appeared.} Indeed, Eq.~(3.50) in~\cite{Burgess:2015nka} subtracts exactly our term $ S_\mathrm{div} $ defined in~\eqref{eq:action_div}.\footnote{However, this subtraction is called a renormalization of the tension in~\cite{Burgess:2015nka}. As discussed above, we disagree with this statement, because a tension would not have the additional metric dependence $ 1 / \sqrt{g_2} $. Consequently, the counter term enters in the Einstein equation~\eqref{eq:einstein_full} differently than $ \mathcal{T}_b $.}
All in all, our EFT analysis is in great agreement with the thorough and detailed UV analysis in~\cite{Burgess:2015nka}, proving once again the usefulness and power of EFT reasoning.


\newpage
\bibliographystyle{utphys}
\bibliography{SLED_BLF_v3}

\end{document}